\documentclass[twocolumn,greek,english,prl,amsmath,superscriptaddress,showpacs]{revtex4}
\usepackage[T1]{fontenc}
\usepackage[iso-8859-7,latin9]{inputenc}
\usepackage{amsbsy}
\usepackage{amssymb}
\usepackage{graphicx}

\makeatletter
\@ifundefined{textcolor}{}
{%
 \definecolor{BLACK}{gray}{0}
 \definecolor{WHITE}{gray}{1}
 \definecolor{RED}{rgb}{1,0,0}
 \definecolor{GREEN}{rgb}{0,1,0}
 \definecolor{BLUE}{rgb}{0,0,1}
 \definecolor{CYAN}{cmyk}{1,0,0,0}
 \definecolor{MAGENTA}{cmyk}{0,1,0,0}
 \definecolor{YELLOW}{cmyk}{0,0,1,0}
}

\makeatother

\usepackage{babel}
\begin{document}

\title{Electronic Raman Scattering On Individual Single Walled Carbon Nanotubes}

\author{Xi Chen}

\address{Department of Physics, The University of Hong Kong, Hong Kong, China}

\author{Bairen Zhu}

\address{Department of Physics, The University of Hong Kong, Hong Kong, China}

\author{Anmin Zhang}

\address{Department of Physics and Beijing Key Laboratory of Opto-electronic
Functional Materials \& Micro-nano Devices, Renmin University of China,
Beijing 100872, People's Republic of China}

\author{Hualing Zeng}

\address{Department of Physics,The Chinese University of Hong Kong, Hong Kong,
China}

\address{Department of Physics, The University of Hong Kong, Hong Kong, China}

\author{Qingming Zhang}

\address{Department of Physics and Beijing Key Laboratory of Opto-electronic
Functional Materials \& Micro-nano Devices, Renmin University of China,
Beijing 100872, People's Republic of China}

\author{Xiaodong Cui}

\address{Department of Physics, The University of Hong Kong, Hong Kong, China}

\email{xdcui@hku.hk }

\begin{abstract}
We report experimental measurements of electronic Raman scattering
under resonant conditions by electrons in individual single-walled
carbon nanotubes (SWNTs). The inelastic Raman scattering at low frequency
range reveals a single particle excitation feature and the dispersion
of electronic structure around the center of Brillouin zone of a semiconducting
SWNT (14, 13) is extracted. 
\end{abstract}


\maketitle
Many experimental techniques for the study of electronic properties
surrender in intrinsic one-dimensional systems particularly single-walled
carbon nanotubes (SWNTs) owing to their low dimension and richness
of geometric structures. Typical examples include the magneto-electric
transport technique and angle-resolved photoemission spectroscopy
which informatively probe the band dispersion (around Fermi level
with electric transport though) and electron's quantum states in
3D and 2D materials, but lose ground in SWNTs. As yet, there lacks
a method capable of directly evaluating the band dispersion in SWNTs.
Resonant Raman spectroscopy has been recognized as one of the most
powerful and popular characterizing technique in SWNT research. SWNTs'
geometric structure could be quantitatively identified at individual
nanotube level with the well established protocols in resonant Raman
spectroscopy\cite{key-1}. Intensive efforts in Raman spectroscopy
focus on the scattering by characteristic phonons as well as the modes
of collective lattice vibrations. And the electronic aspects are implicitly
addressed in Raman study through scattering intensity owing to various
resonant conditions, energy shift due to electron-phonon coupling\cite{key-2}and
spectrum lineshape in presence of Breit-Wigner-Fano (BWF) lineshape
due to phonon-plasmon coupling and electron-electron interactions\cite{key-1,key-3,key-4}.

Recently Farhat et al. reported a mode of electronic Raman scattering
(ERS) from metallic SWNTs, where the Raman shift changes with the
excitation energy and the energy of scattered photon is exactly resonant
with the $M_{ii}$ excitonic transition energy\cite{key-5}. The electric-doping
modulation on the ERS shows that this mode of ERS originates from
the electron-hole excitation by Coulomb exchange at the linear band
of metallic SWNTs. However, the other ERS modes, for example, the
ones arising from single particle and collective elementary excitations
which carry information on the energy and wave-vector dispersions
have not been observed yet. There exist several obstacles toward the
observation of ERS in SWNTs: (i) It is overwhelmed by the substrate
scattering and lattice vibration scattering at finite temperature;
(ii) The ERS originating from single particle and collective excitations
is usually around a few $meV$ in terms of energy, and therefore very
close to the excitation laser. It is technically challenging to distinguish
such low frequency Raman signal from Rayleigh scattering; (iii) Most
resonant Raman scattering on individual SWNTs is collected with confocal-like
micro-Raman setup to maximize the light collecting power, where the
incident light and scattered light are both at normal direction with
respect to the SWNT axis. The momentums of the incident and scattered
photons are orthogonal to the electron momentum dispersion and therefore
the photon and electron elementary excitations are decoupled at normal
incidence due to 1D nature of SWNTs. Here we report the observation
of electronic Raman scattering around $1meV$ from a small suspended
SWNT bundle. We study the low frequency Raman spectra on SWNTs under
the resonant excitation at oblique incident angles. The energy of
the ERS shows linearly proportional to the momentum exchange with
the interacting photons along the SWNT axis. We attribute the ERS
to the single-particle excitation, and the energy-momentum dispersion
of the resonant band could be directly evaluated.

\begin{figure}
\includegraphics[width=9cm]{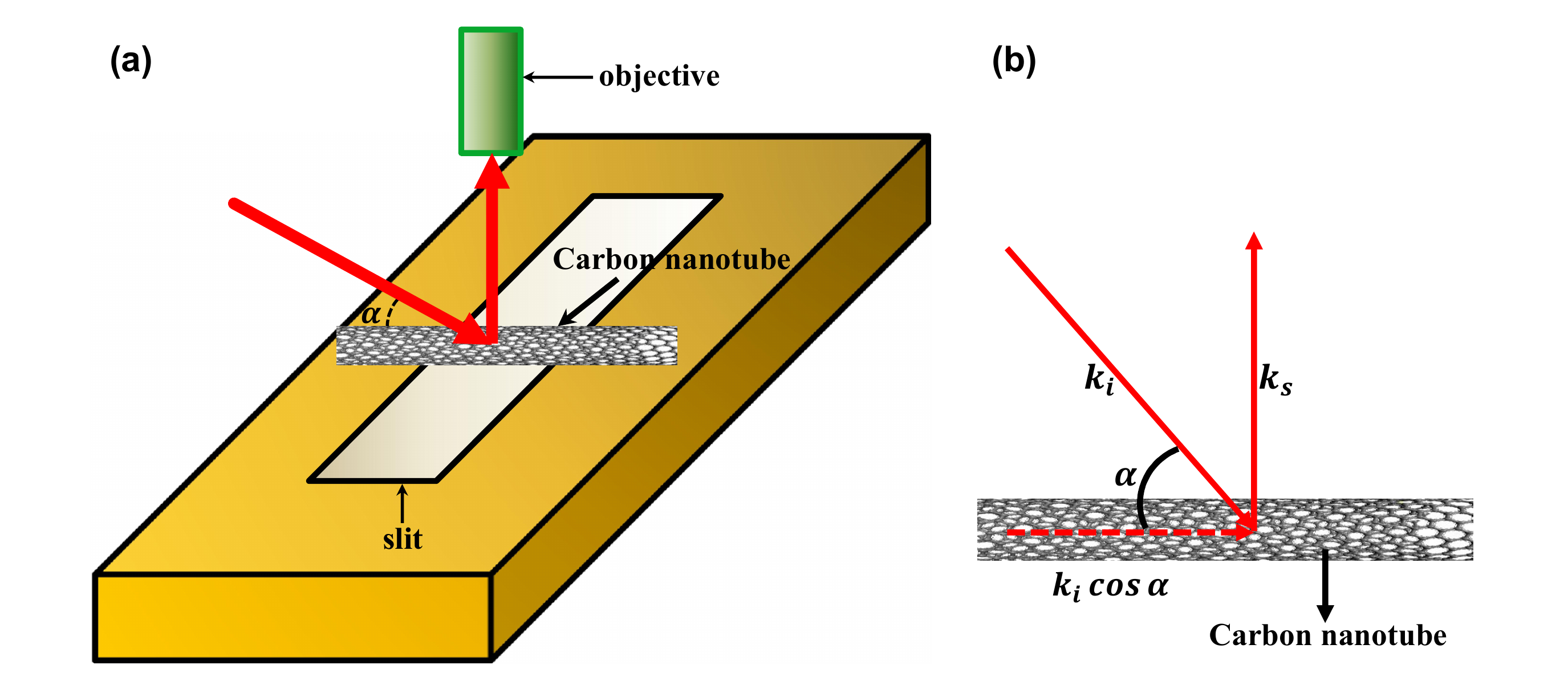}\caption{(a) An illustration of a SWNT suspended on the substrate etched with
open slit and the excitation laser light is shed at an oblique angle
\foreignlanguage{greek}{$\alpha$} against the SWNT tube axis. (b)
Schematic of the scattering geometry with respect to the nanotube
direction.}
\end{figure}

$20\sim30\mu m$ wide $1mm$ long slits on silicon substrates were
fabricated with a standard microelectromechanical (MEMS) process,
including low pressure chemical vapor deposition (LPCVD) silicon nitride
etching mask growth, optical lithography, reactive ion etching (RIE)
and wet etching. The SWNTs were in situ synthesized by chemical vapour
deposition (CVD) across the slits. The catalyst was prepared by selectively
dipping the diluted solution of $FeCl_{3}$ on the silicon substrate
and then by being reduced under $Ar/H_{2}$ 400 SCCM/50 SCCM at $900^{\circ}C$
for $20min$. Individual SWNTs were grown on the substrates in ethanol
vapor with the same gas mixture at $900^{\circ}C$ for $1h$. The
low frequency Raman signals were collected with a $Brag-Grate^{TM}$
notch filter and a single stage monochromator(HR800, Jobin Yvon),
which demonstrates much higher throughput than dual or triple stage
monochromators, with high Rayleigh rejection rate at low frequency
range. The ERS was observed with the resonant laser (HeNe, $633nm$)
at oblique angels against the SWNT axis, as sketched in Figure 1.
Rayleigh scattering spectroscopy was carried out with a supercontinuum
photonic fiber in a similar way as ref \cite{key-6,key-7}.
\begin{figure}
\includegraphics[width=8.7cm]{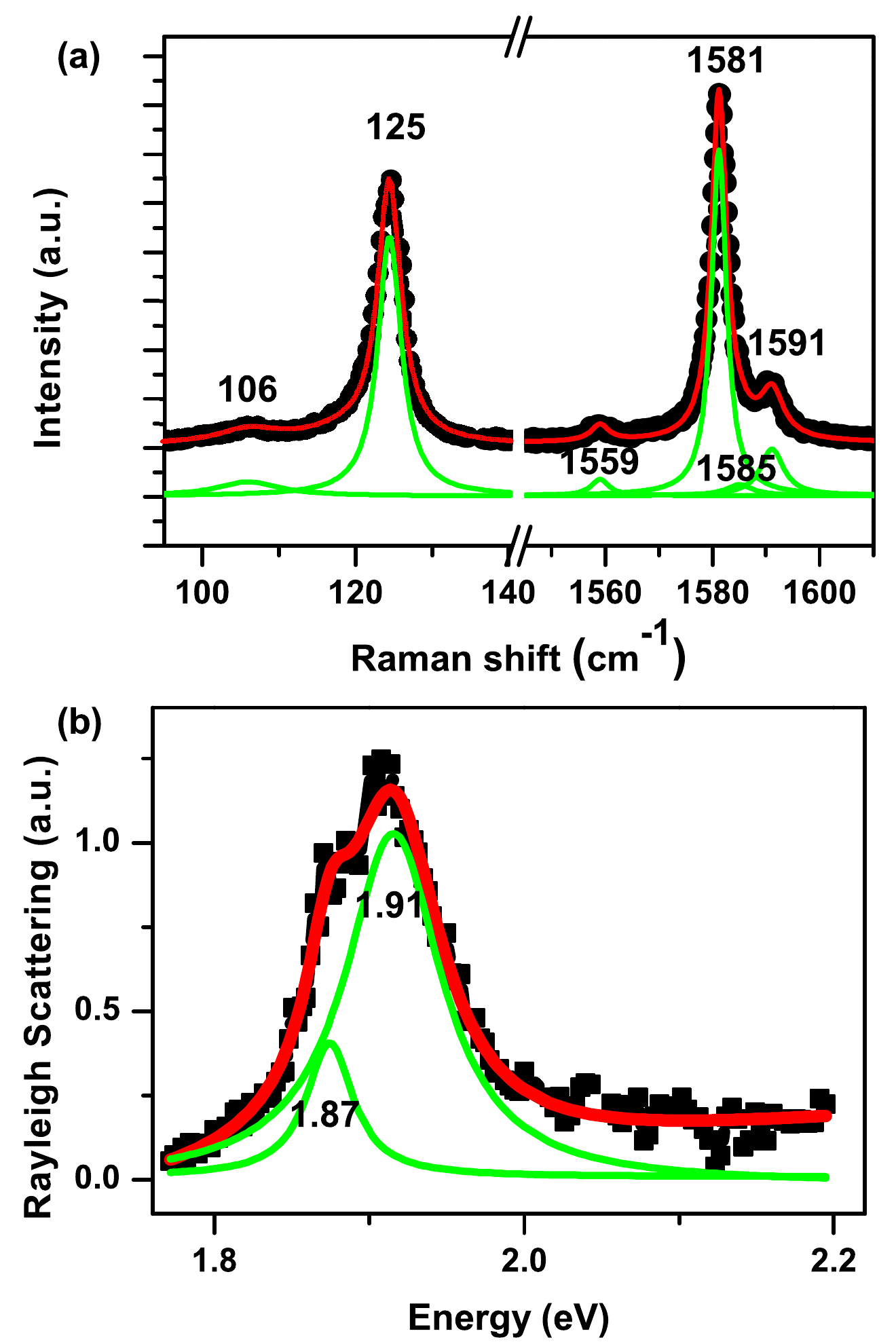}

\caption{(a) The RBM and G-band Raman spectra (black scattered lines), fittings
of sum of multiple Lorentzian peaks (red curves), and separate fittings
(shifted vertically for clarity) of each Lorentzian peaks (green lines)
of the small carbon nanotube bundle. (b) Rayleigh scattering spectrum
of the sample (black scattered line), a general fitting (red curve)
of sum of two peaks, and separate fittings of each peak (green lines)
from the $S_{33}$ transition of SWNT (14, 13) and $S_{44}$ transition
of SWNT (23, 9).}
\end{figure}

Characteristic resonant Raman spectra and Rayleigh scattering spectroscopy
were used to identify the sample as shown in Figure 2. G-band Raman
spectroscopy as shown in Figure 2(a) indicates that our sample is
a small bundle of semiconducting SWNTs with frequency $\omega_{G^{+}}=1591cm^{-1}$
and linewidth $\Gamma_{G^{+}}=6cm^{-1}$, as well as $\omega_{G^{-}}$
at $1559cm^{-1}$, $1581cm^{-1}$ and $1585cm^{-1}$ which can be
tentatively used to estimate the diameter of the tubes\cite{key-8,key-9}.
From the linear relation of $d=228/\omega_{RBM}(nm\cdot cm^{-1})$
between nanotube diameters and the inverse of their Radial Breathing
Mode (RBM) frequencies of suspended SWNTs\cite{key-10}, we further
obtained values of diameter of our SWNTs to be $1.8nm(\omega_{RBM}=125cm^{-1})$
and $2.2nm(\omega_{RBM}=106cm^{-1})$. The interband transition $S_{ii}$
was determined by Rayleigh scattering spectroscopy. As shown in Figure
2(b), Lorentzian fitting $I=\frac{C}{\gamma^{2}\omega^{2}+(\omega^{2}-\omega_{C}^{2})^{2}}$
yields two peaks at $1.91eV$ and $1.87eV$ respectively. According
to the atlas of carbon nanotubes\cite{key-11}with effect of intertube
coupling in bundles considered\cite{key-11-1}, our samples are assigned
to consist of (14, 13) SWNT and (23, 9) SWNT with Rayleigh scattering
peak positions redshifted by $20meV$ and $50meV$ respectively. Values
of diameter of these two SWNTs calculated by $d_{t}=\sqrt{3}a_{CC}(m^{2}+mn+n^{2})^{1/2}/\pi$
( $a_{CC}=0.142nm$ is the nearest-neighbor C-C distance)\cite{key-10}
are consistent with those given by RBM Raman spectroscopy (Figure
2(a)). The peak at $1.91eV$ in Rayleigh scattering spectrum shown
in Figure 2(b) is from the third optical transition ($S_{33}$) of
SWNT with chiral indices (14, 13), while the one at $1.87eV$ corresponds
to $S_{44}$ of nanotube (23, 9).

\begin{figure}
\includegraphics[width=9.5cm]{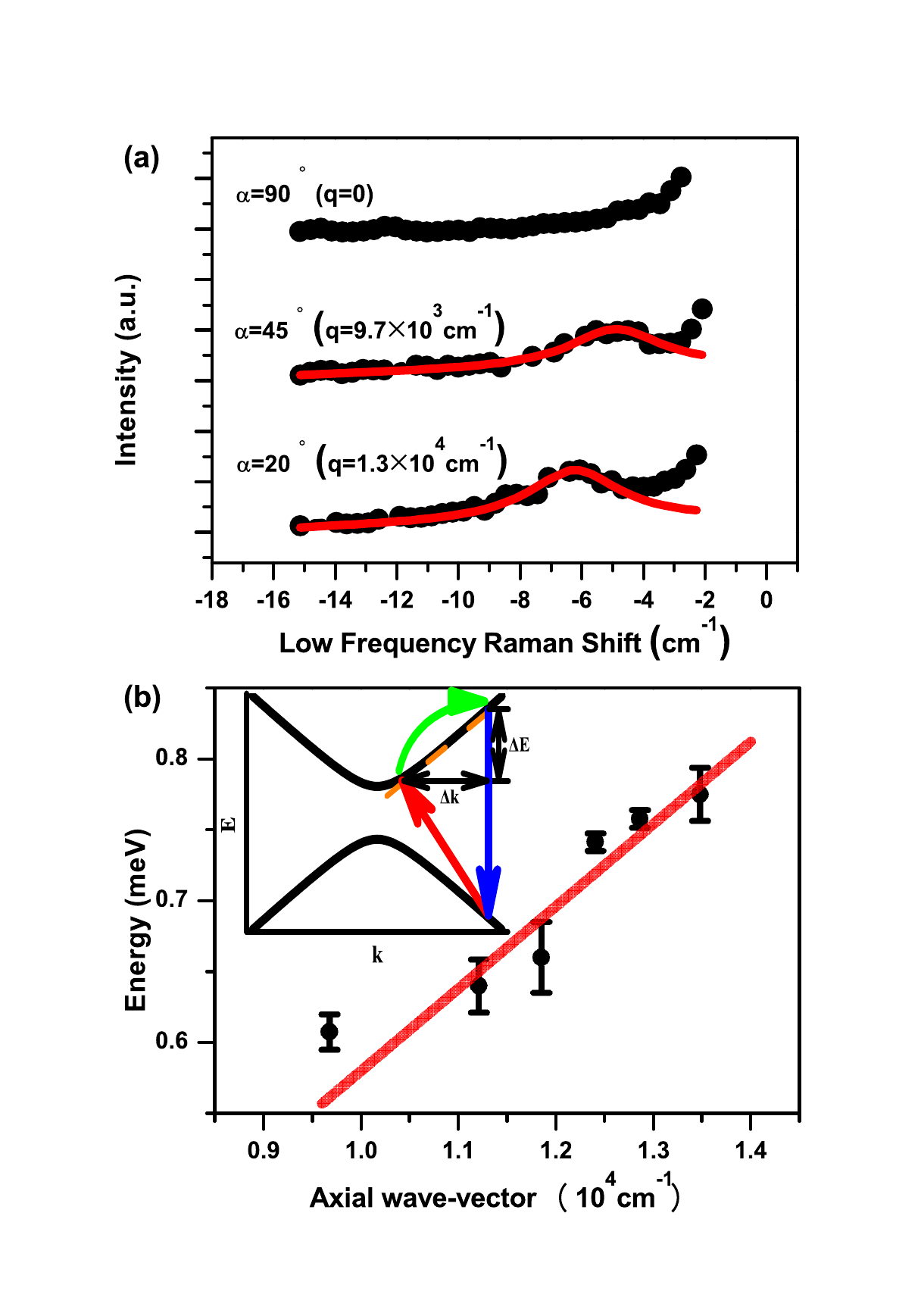}

\caption{(a) Representative Raman scattering (black scattered lines) at incident
angle of $90^{\circ}$, $45^{\circ}$and $20^{\circ}$(q denotes the
corresponding magnitude of projection of wave-vector along the SWNT
axis at certain angle). The red lines following a Lorentzian lineshape
are for highlight only. (b) Raman shift (in unit of energy) as a function
of axial wave-vector (black scattered dots with error bars) exhibiting
linear variations and the linear fitting (red straight line) of the
data . Inset: schematic picture of the single particle excitation
(SPE) model.}
\end{figure}

Figure 3 shows the representative low frequency Raman scattering from
the SWNT under resonant excitations at oblique incident angles. A
small bump (Figure 3(a)) gradually rises from none at normal incidence
to a few wave number (around $1meV$) at oblique incidence. The bump
shows blue-shifted with the decrease of incident angle. The corresponding
energy has a linear dependence on $k_{i}cos\alpha$, the projection
of the incident photon wave-vector along the SWNT axis. To our knowledge,
the lowest frequency of phonon mode at $\Gamma$-point was in the
range of $10\sim12cm^{-1}$\cite{key-12}, bigger than our Raman results
with frequency ranging from $4.8$ to $6.2cm^{-1}$. Besides, slope
of acoustic phonon dispersion near $\Gamma$-point can be estimated
to be $4\backsim15\times10^{-7}meV\cdot cm$\cite{key-13}, which
is two orders of magnitude smaller than our results. Therefore, possible
origins associated with phonon dispersion can be ruled out. The position
of the small bump is independent of the excitation intensity in the
range of $1\sim14mW$. Thus the bump unlikely originates from the
Plasmon or other collective modes, as the corresponding dispersion
is a monotonic function of effective carrier density and the observed
features are independent of the excitation intensity. We attribute
the low frequency mode to single particle excitation (SPE) of SWNT
(14, 13), as illustrated in the inset of Figure 3(b). In the experimental
setup, the incident angle concludes the photon's momentum projection
$k_{i}cos\alpha$ along the SWNT axis and therefore determines momentum
transfer between the incident and scattered photons. As the requirements
of energy conservation and momentum conservation along the SWNT axial
direction, the energy difference between the incident and scattered
photons exactly reflects the electronic band dispersion at resonant
energy. Subsequently, the Raman frequency increases with the increase
of wave-vector transfer. As the magnitude of the wave-vector transfer
is quite small $(q\ll1/a)$ , the quotient of $\frac{\triangle E}{\triangle k}$
well presents the slope of the electronic band dispersion. Meanwhile
the joint density of states of SWNTs follows $\frac{1}{\sqrt{E-E_{ii}}}$
as a result of quasi 1D confinements, and consequently the electronic
Raman process only occurs upon the resonance, namely around the inter-subband
edge. Therefore, from the relation between Raman energy and the wave-vector
transfer as plotted in Figure 3(b), we can estimate the slope of electronic
energy band at subband edge. A general linear behavior is also fitted
in Figure 3(b), which has an intercept of zero as expected from the
band dispersion. Axial wave-vector $\triangle k$ ranging from $9673cm^{-1}$
to $13473cm^{-1}$ is obtained in our experiments, and the corresponding
slope is calculated to be in the range of $5.6\backsim6.3\times10^{-5}meV\cdot cm$.

To demonstrate the reliability of our experimental results, we apply
the empirical formula\cite{key-11}below as effective dispersion relation
to estimate the slope of electronic band dispersion:

$E_{p}(\mathbf{\boldsymbol{\mathit{k}}})=2\hbar v_{F}(p)\times\mathbf{\mathit{k+\beta\times k^{2}+\eta(p)\times k^{2}cos(\mathit{\mathit{\mathit{3\theta}}})}}$,
with $v_{F}=1.221\times10^{6}m\cdot s^{-1}$,$\beta=-0.173eV\cdot nm^{2}$,$\eta=0.058eV\cdot nm^{2}$
for $S_{33}$ transition.

Given energy of the excitation light of our experiments being $1.959eV$
(wavelength at $633nm$), we can calculate the magnitude of corresponding
wave-vector on the electronic energy band of SWNT (14, 13) from the
dispersion to be $1.446\times10^{7}cm^{-1}$, and the slope at this
point to be about $1.1\times10^{-4}meV\cdot cm$. The calculated result
qualitatively shows agreement with what we observed in the experiments.
On the other hand, if we follow the model of exciton Kataura plots
with environment corrections\cite{key-14,key-15}, our sample would
be assigned to a bundle composed of a (23, 1) nanotube and a (18,
11) nanotube. In T. Ando's theory\cite{key-16}, the energy dispersion
is described as $E(i,k)=\gamma\sqrt{\kappa_{\nu}(i)^{2}+k^{2}}$,
where $\kappa_{\nu}(n)=\frac{2\pi}{L}(n-\frac{\nu}{3}),\gamma=\frac{\sqrt{3}}{2}a\gamma_{0}$,
$a=0.246nm$, $\gamma_{0}$ refers to the transfer integral between
nearest-neighbor carbon atoms and is assumed to be $2.9eV$\cite{key-17}.
$S_{33}$ transition of SWNT (23, 1) is now associated with peak at
$1.91eV$, so the slope $\frac{\partial E(3,k)}{\partial k}$ is calculated
to be in the order of $1\times10^{-5}meV\cdot cm$. This is qualitatively
consistent with our experimental results, which provides a strong
support for our explanation of SPE energy dispersion picture.

In summary, we report a low-frequency Raman mode from SWNTs under
resonant excitations at oblique incidence. The corresponding Raman
mode shows a linear dependence on the momentum transfer along the
SWNT. We attribute the Raman mode to the scattering from electronic
SPE. The slope of electronic band dispersion at the subband edge is
measured around $5.6\backsim6.3\times10^{-5}meV\cdot cm$ on SWNT
with structural index (14, 13). 
\begin{acknowledgments}
The project was supported by the Hong Kong research grant councel
under HKU 701810P, SRT on New Materials of University of Hong Kong,
the Ministry of Science and Technology of China (973 projects: 2011CBA00112
and 2012CB921701) and NSF of China (Grant No.: 11034012 \& 11174367).. \end{acknowledgments}

\end{document}